\documentclass[useAMS,usenatbib]{mn2e}
\usepackage{graphicx,amssymb,natbib}
\usepackage{longtable}

\title[Search for variable stars]
{Photometric search for variable stars in young open cluster Berkeley 59}

\author[Sneh Lata et al. ]
       {Sneh Lata \thanks{E-mail: sneh@aries.res.in}, A. K. Pandey, Maheswar G., Soumen Mondal and Brijesh Kumar \\ 
        Aryabhatta Research Institute of Observational Sciences, Manora Peak, Nainital 263129, Uttarakhand, India}
\date{Accepted ---------.
      Received ---------;
      }

\pagerange{\pageref{firstpage}--\pageref{lastpage}}

\def\LaTeX{L\kern-.36em\raise.3ex\hbox{a}\kern-.15em
    T\kern-.1667em\lower.7ex\hbox{E}\kern-.125emX}

\begin{document}

\label{firstpage}

\maketitle

\label{firstpage}
\begin{abstract}

We present time-series photometry of stars located in the extremely young open cluster Berkeley 59. 
Using the 1.04 m telescope at ARIES, Nainital, we have identified 42 variables in a field of $\sim13^{'}\times13^{'}$
 around the cluster.  
The probable members of the cluster are identified using $(V, V-I)$ colour-magnitude diagram and $(J-H, H-K)$ colour-colour
diagram.
Thirty one variables are found to be pre-main sequence stars associated with the cluster. 
The ages and masses of pre-main sequence stars are derived from colour-magnitude diagram by fitting theoretical models to the
observed data points.
The ages of the majority of the probable pre-main sequence variable candidates range from 1 to 5 Myrs.
The masses of these
 pre-main sequence variable stars are found to be in the range of $\sim$0.3 to $\sim$3.5 $M_{\odot}$ and these could be T Tauri stars. 
The present statistics reveal that about 90\% T Tauri stars have periods $<$ 15 days.
The classical T Tauri stars
are found to have larger amplitude in comparison to the weak line T Tauri stars.
There is an indication that the amplitude decreases with increase of the mass, which could be due to the dispersal of disk of relatively massive stars. 
\end{abstract}

\begin {keywords} 
Open  cluster:  Berkeley 59  --
colour--magnitude diagram: Variables-pre-main sequence stars
\end {keywords}

\section{Introduction}
Variable stars are useful tools to improve our understanding on stellar evolution and structure of stars.
Star clusters are unique laboratories to study the stellar evolution as star clusters
provide a sample of stars having same age, distance, initial composition and spanning range of masses. 
Additionally, the light curves of several stars can be produced simultaneously to know
the real variable stars among them.
Studies of variable stars in star clusters are carried out since several years (Herbst et al. 1994; Edwards et al. 1993; Oliveira \& Casali 2008; Grankin et al. 2008 ). 
Star clusters have different types of variable stars as stars show variability at various stages
of their evolutionary phases. 

In the present study, we made a search for variable stars in an extremely young open cluster Berkeley 59 (Be 59).
 Be 59 is a young cluster and
 it contains a significant numbers of pre-main sequence (PMS) stars i.e, stars evolving from their birthline
to the main-sequence (MS).
This cluster ($\alpha_{2000}$ = 00${^h}$ 02${^m}$ 13${^s}$, $\delta_{2000}$ = +67$^{o}$ 25$^{'}$ 11$^{"}$, l = 118$^{o}$.22, b = 5$^{o}$.00) is located at the centre of OB4 stellar association surrounded by HII region, Sharpless region S171 (Yang \& Fukui 1992).
 Previous studies have revealed that Be 59 consists of many PMS stars having age around 2 Myr (Pandey et al. 2008).
Be 59 was first studied photographically by Blanco \& Williams (1959). They concluded that stars of spectral types O7 to B5 are still surrounded by a part of the parent molecular cloud.
A multiwavelength study on Be 59 was presented by Pandey et al. (2008, hereafter P08). They found that this cluster
is located at a distance of 1.0 kpc and having a variable extinction of $E(B-V)$=1.4-1.8 mag.
They also found that around 32\% of the H${\alpha}$ emission stars in Be 59 exhibit NIR excess indicating that they still have their inner disk. During a campaign of photometric monitoring,
Majaess et al. (2008) presented the nature of the three eclipsing
systems (BD+66$^{o}$1673, 2MASS 00104558+6127556 and 2MSASS 19064659+4401458) in
the field of Be 59 using $BV$ photometry and spectroscopic observations.
From the above studies it is clear that Be 59 contains high, intermediate and low mass PMS stars.

PMS objects are, in general, classified into T Tauri stars (TTSs) and Herbig Ae/Be stars.
The TTSs are newly formed low-mass stars ($\lesssim$3M${_\odot}$) which are contracting towards the MS
 (Herbig 1977), while stars of intermediate mass ($\gtrsim$3-10 M$_{\odot}$) are known as Herbig Ae/Be stars (Herbig 1960; Finkenzeller \& Mundt 1984; Strom et al. 1972).
Herbig Ae/Be stars are either contracting towards MS or they have reached
the MS.
The TTSs can be classified further as Weak line TTSs (WTTSs) and Classical TTSs (CTTSs) on the basis of the strength of the H${\alpha}$ emission lines (Strom et al. 1989).
The strength of the emission line is measured by its equivalent width (EW).
The WTTSs exhibit a weak, narrow H$\alpha$ (EW$\le10\AA$) in emission with no or little infrared excess, while CTTSs generally display strong H${\alpha}$
emission line (EW\(>\)10$\AA$), large ultraviolet and infrared excess.
The large  H$\alpha$ emission (EW\(>\)10$\AA$) is believed to be a result of the increased mass accretion onto the star (e.g. Cabrit et al. 1990; Calvet \& Hartmann 1992). 
 It is found that PMS stars show 
strong variability both in photometric and spectroscopic (van den Ancker et al. 1998; 
Catala et al. 1999) observations. This could be due to the presence of circumstellar obscuration and spots present on the photosphere. 
In fact variability is a defining characteristics of TTSs (Herbst et al. 1994; Scholz et al. 2009 and references therein).
Both WTTSs and CTTSs show variation in their brightness. 
These variations are found to occur at all wavelengths, from X-ray to infrared.
Variability time scale of TTSs ranges from few minutes to years (Appenzeller \& Mundt 1989). 
The photometric variations are believed to originate from several mechanisms like circumstellar
gas and dust (remnant of parent molecular cloud), accretion and magnetic field (Herbst et al. 1994).
The variations in the brightness of TTSs are most probably due to the presence of spots (cool or hot) at stellar surface and circumstellar disk. 

The cool spots on the surface of the stars are produced by the emergence of stellar magnetic fields on the photosphere, and is thus an indicator of magnetic activity.
As these spots are present on the photosphere, they rotate with the star. If these spots are symmetrically distributed over the photosphere, periodic variations are seen in the light curves of the stars.
Spot configurations remain constant over time-scales of several days, but undergo changes on time-scales of weeks (Hussain 2002). 
The cool spots on the photosphere of star are responsible for brightness variation in WTTSs, and these objects are found to be fast rotators as they have either thin or no circumstellar disk.
Therefore the variability in these objects may be mainly due to the brightness variations on the photosphere modulated
by the stellar rotation (Messina et al. 2010).

The hot spots on the surface of young stars are supposed to be formed on the surface by infalling gas and are thus a direct consequence of accretion
(Lynden-Bell \& Pringle 1974; Koenigl 1991; Shu et al. 1994). 
Irregular or non periodic variations are produced because of changes in the accretion rate. The time scales of varying brightness
range from hours to years. 
 The hot spots cover a smaller fraction of the stellar surface but
higher temperature causes larger brightness variations (Carpenter et al. 2001).
Thus, accreting  CTTSs which are surrounded by circumstellar accretion disk and possess hot spots produced by accreting material from disk on to the star show complex behaviour in their optical and near infrared (NIR) light curves (Scholz et al. 2009).

\begin{figure*}
	\includegraphics[width=14cm]{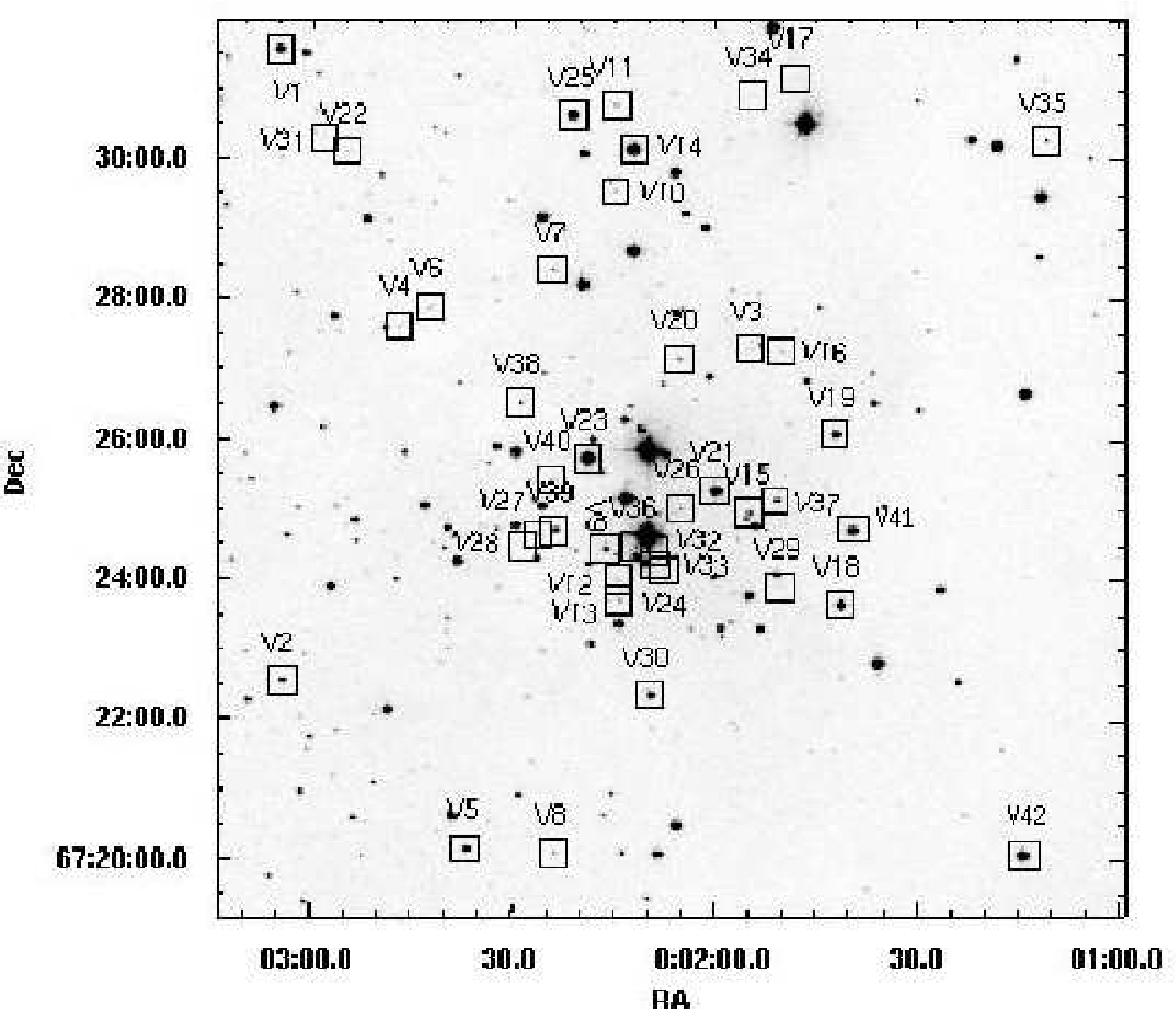}
\caption{ The observed CCD image of Be 59 cluster region. The squares show the location of candidate variables identified in the present work.}
\end{figure*}

The Herbig Ae/Be stars also show variability
as they move across the instability region in the Hertzsprung-Russell (HR) diagram on their way to the MS. 
The right combination of mass, temperature and luminosity of these stars make them to pulsate.
Several PMS pulsators have been detected until now
(Breger 1972; Kurtz \& Marang 1995; Marconi et al. 2000;
Donati et al. 1997; Zwintz et al. 2005, 2009; Zwintz \& Weiss 2006).
Catala (2003) presented detailed review on the PMS pulsations and expected that PMS stars could pulsate as $\delta$ Scuti stars, however it is necessary to 
confirm their PMS nature because these stars are found above the MS where it is very difficult to say whether they are PMS or post MS stars. 
But in cluster environment where the age of the cluster is constrained well, the stars found towards the cluster region could be PMS stars.

A search of photometric variables in young open cluster Be 59, which is found to have a rich population of PMS stars with a large mass range could give us a better insight of the variability
time-scales, the amplitude of brightness variation etc. With this aim, we observed the cluster Be 59 at different epochs from 2006 to 2010.
The paper is organised in the following manner. In Section 2, we describe the observations and data
reduction techniques. We discuss membership of variable stars on the basis of photometric observations in Section 3. In Section 4, we determine periods of variable stars. We discuss variability characteristics as well as period and amplitude distribution of variable stars in Section 5, while in Section 6 we conclude our studies. 

\begin{table}
\centering
\caption{Log of optical photometric CCD observations. `Exp.' and `n' refer to exposure time and number of frames respectively. \label{tab:obsLog}}
\begin{tabular}{llcc}
\hline
date&
Object&
{\it V}&
{\it I} \\
&
&
(n$\times$ Exp.)&
(n$\times$ Exp.) \\
\hline
21 Dec 2006 & Be 59&3$\times$300s, 700s, 700s, 500s& 4$\times$300\\
22 Dec 2006 & Be 59&5$\times$700s&-\\
23 Dec 2006 & Be 59&2$\times$700s&-\\
24 Dec 2006 & Be 59&2$\times$700s&3$\times$300s\\
24 Dec 2006 & SA 98&3$\times$200s&3$\times$200s\\
25 Dec 2006 & Be 59&1$\times$700s&-\\
05 Dec 2007 & Be 59&43$\times$50s&-\\
25 Oct 2008 & Be 59&58$\times$50s&-\\
29 Oct 2008 & Be 59&3$\times$50s&-\\
21 Nov 2008 & Be 59&3$\times$50s&-\\
11 Oct 2009 & Be 59&60$\times$300s&-\\
12 Oct 2009 & Be 59&3$\times$300s&- \\
13 Oct 2009 & Be 59&2$\times$300s &-\\
14 Oct 2009 & Be 59&2$\times$300s&- \\
15 Oct 2009 & Be 59&2$\times$300s &-\\
27 Oct 2010 & Be 59&58$\times$200s &-\\
28 Oct 2010 & Be 59&3$\times$200s &-\\
30 Oct 2010 & Be 59&2$\times$200s &-\\
31 Oct 2010 & Be 59&2$\times$200s &-\\
\hline
\end{tabular}
\end{table}

\section{Observations and Data Reduction}
Our observing campaign of Be 59 region
covers
18 nights from 2006, December 21 to 2010, October 31. 
Fig. 1 shows the observed field of Be 59.
The observations in $V$ band were obtained using 2k$\times$2k CCD camera
attached to the 104-cm ARIES telescope. As each pixel covers 0.38 arcsec over sky,
the field of view is $\approx$ 13 arcmin on each side. On each night, at least two 
frames of target field in $V$ band were secured to cover long period 
variables. The observations of Be 59 consist of total 257 CCD images.
 In
addition, bias and twilight
frames were also taken on each night. In order to improve the signal to noise ratio the observations were taken in 2$\times$2 pixel binning mode.
Log of the observations is given in Table 1.
The preprocessing of the CCD images
was done by IRAF which includes bias subtraction, flat fielding and cosmic ray removal.
The instrumental magnitude of each star was obtained by DAOPHOT software package given by Stetson (1987).
To get the magnitude of each star both the aperture and PSF photometry have been done because the PSF photometry
yields better results in crowded region.
The standardisation of the observations was
done by observing the standard field SA 98 (Landolt 1992). 
Instrumental magnitudes were converted to standard
magnitudes using the following transformations equations \\
$$v = V+(4.448\pm0.004)-(0.037\pm0.006)(V-I) + 0.15*Q $$
$$i= I+(4.767\pm0.008)-(0.04\pm0.005)(V-I)+0.05*Q $$
\noindent where $v$ and $i$
are the  instrumental magnitudes and $Q$ is the airmass.

\begin{figure}
\includegraphics[width=8cm]{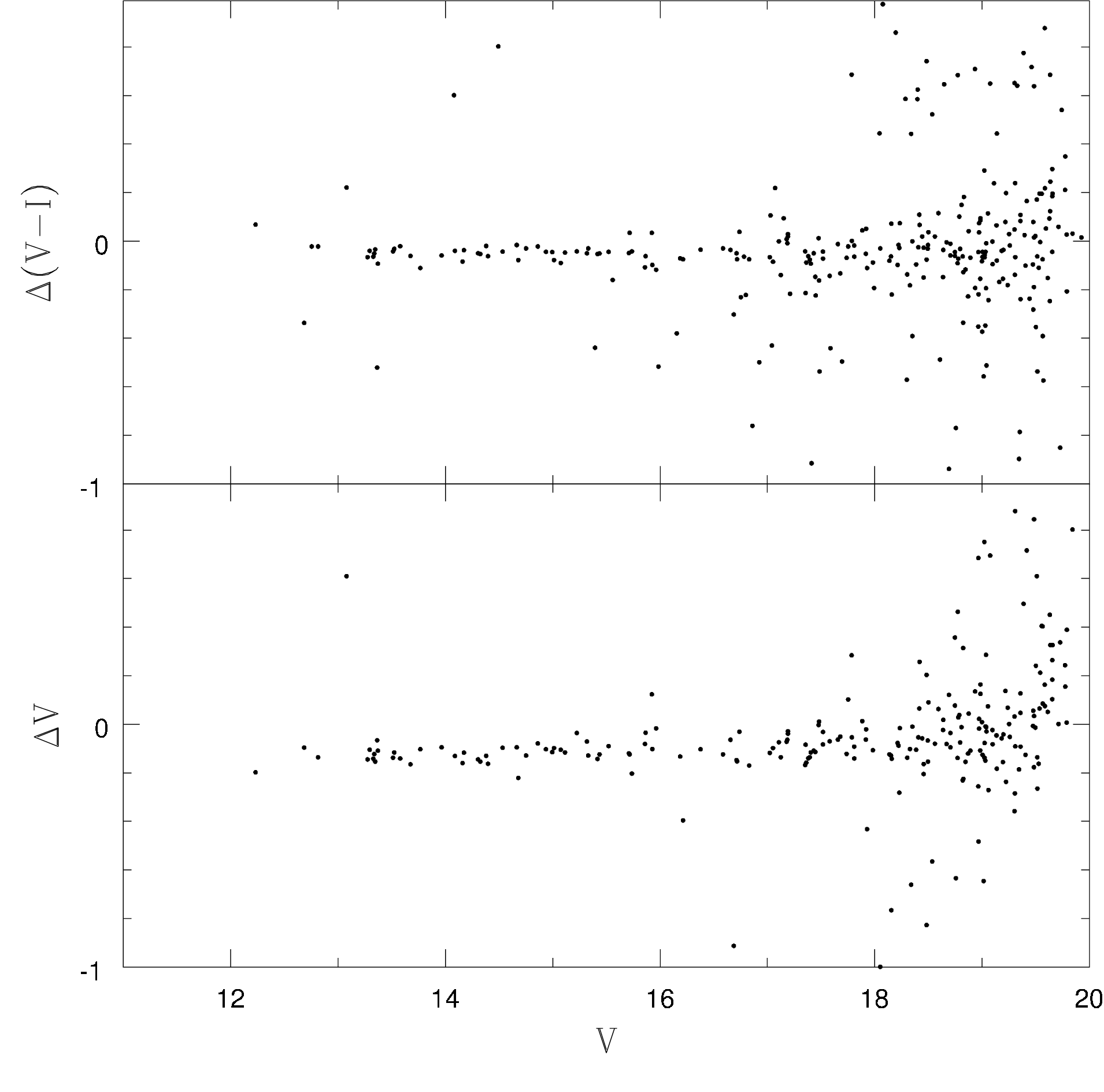}
\caption{Comparison of the present photometry and photometry given by Pandey et al. (2008).}
\end{figure}

\begin{figure}
\includegraphics[width=8cm]{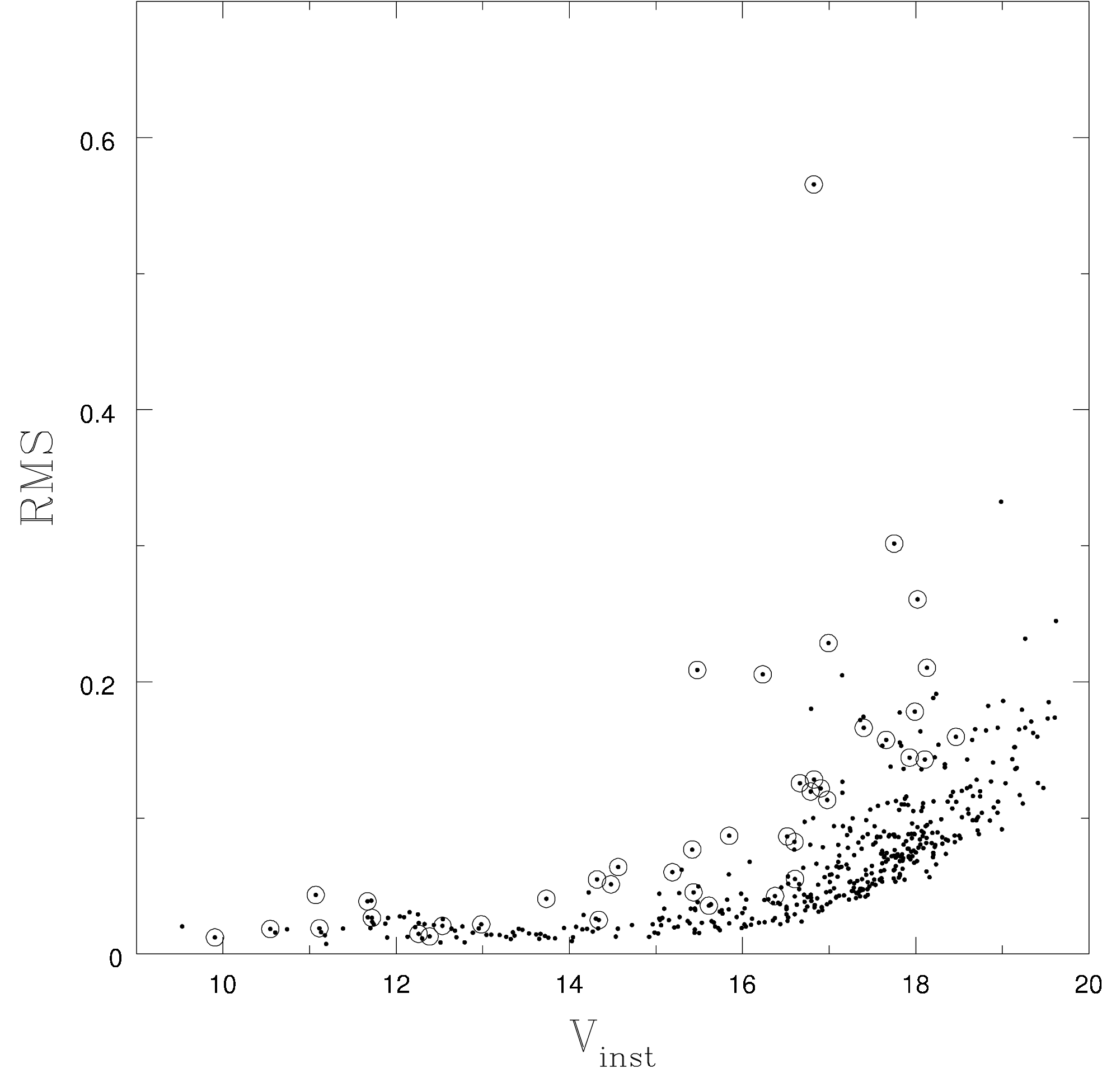}
\caption{RMS for each star as a function of brightness. Open circles represent
candidate variables identified in the present work. Some stars showing large RMS are not considered variable stars as they are located near the edge of the CCD. }
\end{figure}

\begin{figure}
\includegraphics[width=8cm]{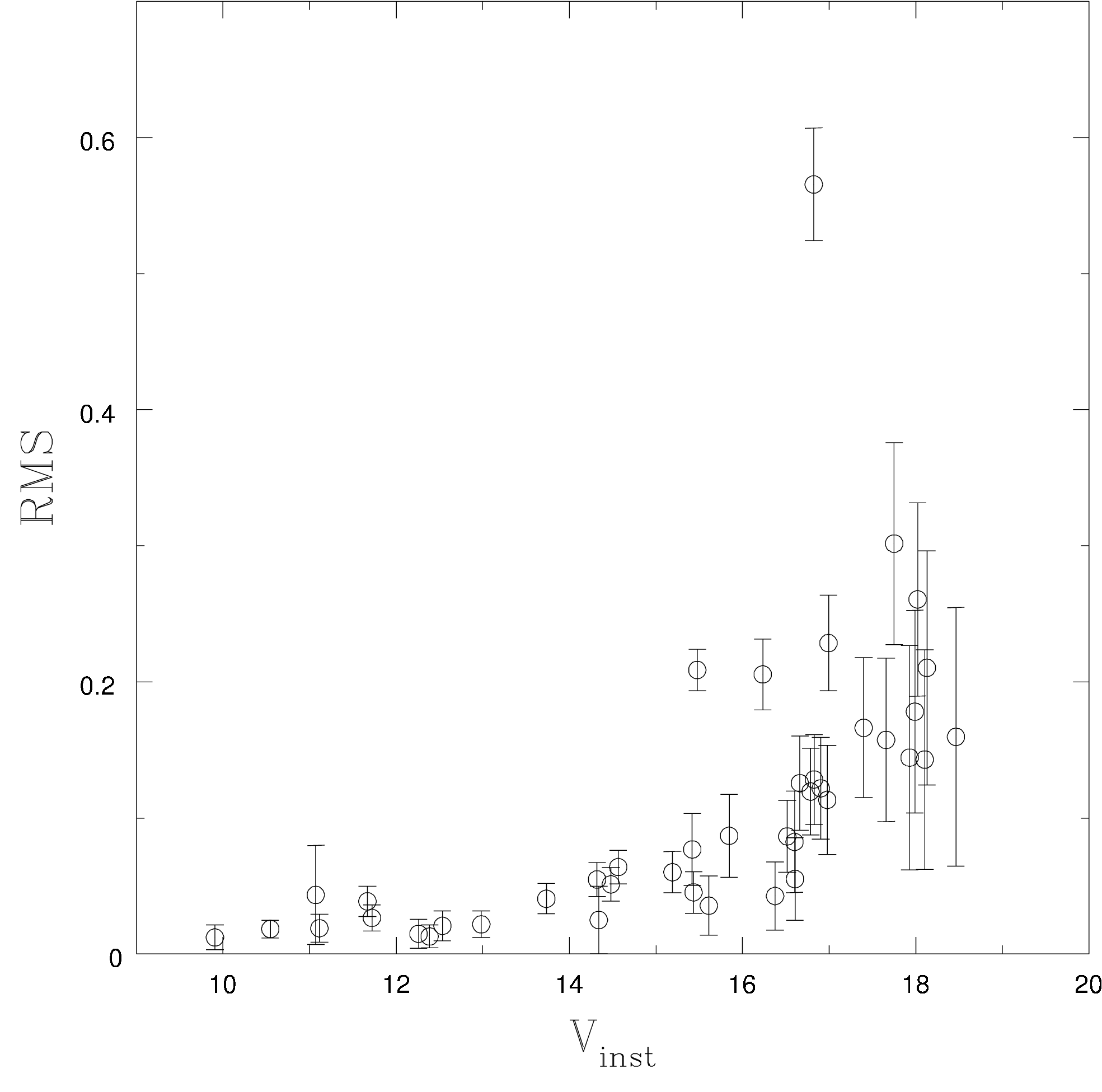}
\caption{The RMS along with mean photometric errors of the detected variable stars as a function of instrumental
magnitude.}
\end{figure}

\subsection{Comparison with Previous Photometry}
A comparison of the present photometry with that of P08 yields 342 common stars.
Fig. 2 plots difference $\Delta$ (present data$-$literature data) as a function of $V$ magnitude.
 The comparison indicates a decreasing trend in $\Delta V$ values. The $\Delta V$ is $\approx$ $-0.13$ mag upto $V$$\approx$14 mag which decreases
to $\approx$ $-0.08$ mag upto $V$$\approx$18 mag.
Whereas present $(V-I)$ colours are comparable to those given by P08.

\subsection{Identification of Variables}
The DAOMATCH (Stetson 1992) routine of DAOPHOT was used to find the translation, rotation and scaling solutions between different 
photometry files, whereas DAOMASTER (Stetson 1992) matches the point sources.
In order to remove frame-to-frame flux variation of stars
due to airmass and exposure time, we used DAOMASTER programme
to get the corrected magnitude.
This task makes the mean flux level of each frame equal to the
reference frame by an additive constant.
Present observations have 257 frames and each frame corresponds to one photometry file.
DAOMASTER cross identified 472 stars in different photometry files and listed their
corrected magnitudes in .cor file, which was
used for subsequent tasks.
The mean value of the magnitude and square root of the variance (RMS) of the data were estimated using the observations 
 of each star. The RMS as a function of instrumental $V$ magnitude ($V_{inst}$) is shown in Fig. 3, which
indicates that the majority of the stars follow an expected trend i.e., S/N ratio decreases as stars become fainter.
However, a few stars do not follow the normal trend and exhibit relatively large scatter. These could be either due
to the large photometric errors or due to the variable nature of the stars. We considered a star as variable if its RMS is
greater than 3 times of the mean RMS of that bin.
Forty five candidate variables were thus identified on the basis of above
mentioned criterion.
On the basis of a careful inspection of light curves and the location of the sources on the CCD frames, we selected a sample of 37 stars to study their nature. A few stars were rejected as they were lying near the edge at the CCD. The mentioned technique, however, could not detect low amplitude variables and 5 variables namely V14, V18, V21, V23 and V19 were 
detected manually by inspecting their light curves. Thus we have a sample of 42 variables. Their identification numbers, coordinates and photometric data
are given in Table 2.
The RMS along with mean photometric errors of detected variables as a function of instrumental magnitude are shown in
Fig. 4. The photometric error is found to be $\sim$ 0.010 mag at $V_{inst}$ $\le$ 14 mag, whereas the value of photometric error increases to $\sim$0.08 mag at $\sim$18 mag.

Out of three variable stars reported by Majaess et al. (2008),  
only one variable star (BD+66$^{o}$1673) 
is 
located in the present observed field of Be 59. Unfortunately, this star was  
saturated in our observations.

\begin{figure}
\includegraphics[width=8cm]{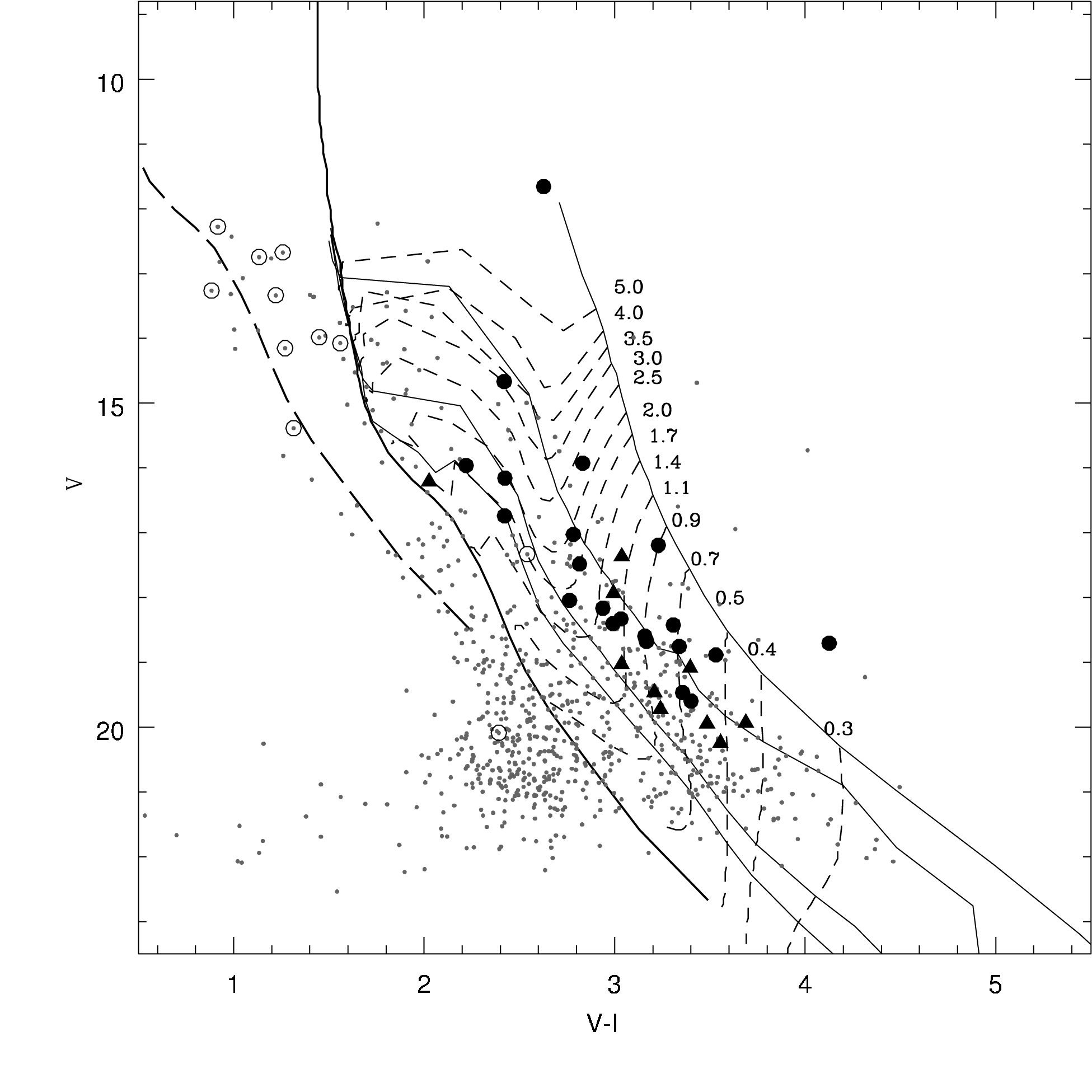}
\caption{$V, V-I$ colour-magnitude diagram of Be 59. The ZAMS by Girardi et al. (2002)
 and PMS isochrones for
0.1, 1, 5, 10 Myrs by Siess et al. (2000) are also shown.
The dashed curves show PMS evolutionary tracks for stars of different masses.
The isochrones and evolutionary tracks  are corrected for the cluster distance and $E(B-V)=1.45$ mag.
The filled circles and triangles represent probable WTTSs and CTTSs respectively (cf. section
3.2). The thick dashed is ZAMS by Girardi et al. (2002) corrected
for a distance of 470 pc and reddening  $E(B-V)=0.40$ mag.
The open circles represent foreground stars.
}
\end{figure}

\begin{figure}
\includegraphics[width=8cm]{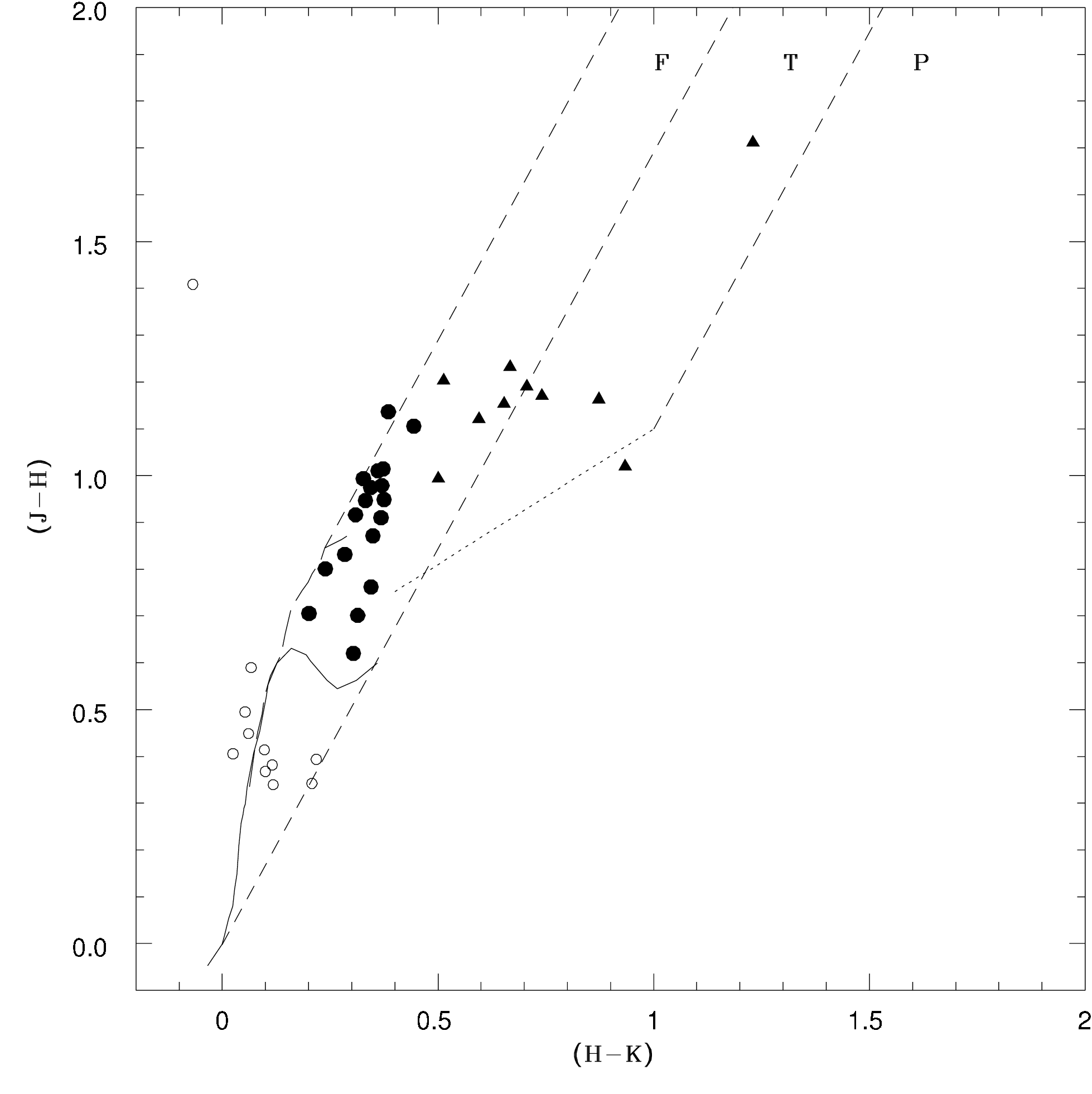}
\caption{($J-H, H-K$) colour-colour diagram for stars lying in the 
13$\times$13 arcmin$^2$ field of Be 59. $JHK$ data have been taken from 2MASS catalogue. The symbols are same as in Fig. 5. 
The sequences for dwarfs (solid) and giants (long dashed) are from Bessel \& Brett (1988). 
The dotted line represents the intrinsic locus of CTTSs (Meyer et al. 1997).
Dashed lines represent the reddening vectors (Cohen et al. 1981).}
\end{figure}


\begin{table*}
\caption{Basic parameters of 42 variables in open cluster Be 59. H${\alpha}$ sources are taken from Pandey et al. (2008). \label{tab:basPar}}
\tiny
\begin{tabular}{llccclcllllll}
\hline
ID&
$\alpha_{2000}$&
$\delta_{2000}$&
$V$&
$(V-I)$&
$J$&
$H$&
$K$&
Age&
Mass&
Amp.&
Period&
Object Type\\
&
&
&
mag&
mag&
mag&
mag&
mag&
Myrs&
$M_{\odot}$&
mag&
day&
\\
\hline
V1& 0.77192 & 67.52431&   13.984&1.449& 11.519&    11.197&     11.045&-    &  -  &          0.023&  0.225  &FGMS$^{1}$  \\
V2&0.76846  & 67.37411&   16.154&2.424& 12.038&    11.318&     11.083& 5.00& 2.50&          0.039&  0.539  &PMS  \\
V3&0.47883  & 67.45319&   19.724&3.241& 15.612&    13.799&     12.535& 3.50& 0.80&          0.855&  89.800   &PMS, IR excess, CTTS  \\
V4&0.69671  & 67.45858&   19.082&3.396&  13.420&    12.128&     11.427& 1.00& 0.65&         0.272&  0.652    &PMS, IR excess, CTTS  \\
V5&0.65492  & 67.33436&   15.386&1.315& 13.118&    12.627&     12.540&-    &  -  &          0.041&  0.508    &FGMS  \\
V6&0.67825  & 67.46291&   18.332&3.033& 13.115&    12.103&     11.725& 1.50& 1.10&          0.066&  0.527    &PMS  \\
V7&0.60192  & 67.47219&   17.366&3.037& 12.066&    10.895&     10.266& 0.50& 1.10&          0.270&  6.586    &PMS, $H{_\alpha}$, IR excess, CTTS \\
V8&0.60004  & 67.33353&   18.045&2.763& 13.244&    12.463&     12.084& 4.00& 1.85&          0.067&  1.775    &PMS \\
V9&0.56908  & 67.40567&   15.925&2.832 &11.047&    10.147&      9.764& 0.15& 2.50&          0.030&  0.974  &PMS \\
V10&0.56275 & 67.49089&   19.024&3.036& 13.175&    11.958&     11.051& 4.00& 1.10&          1.122&  1.102,58.753    &PMS, $H{_\alpha}$, IR excess, CTTS \\
V11&0.56254 &  67.51144&  18.402&2.991 &13.141&    12.157&     11.748& 2.00& 1.25&          0.122&  0.441    &PMS  \\
V12&0.56121 &  67.39845&  18.753&3.339& 12.968&    11.917&     11.522& 0.80& 0.70&          0.273&  5.671    &PMS \\
V13&0.56029 &  67.39361&  17.931&2.994& 12.486&    11.261&     10.486& 1.00& 1.25&          0.290&  1.212    &PMS, IR excess, CTTS \\
V14&0.55221 &  67.50259&  12.427&0.947& 10.693&    10.340&     10.206& -   & -   &          0.016&  0.158    &FGMS \\
V15&0.47992 &  67.41450&  16.210&2.026& 12.733&    11.672&     10.704& 20.00& 2.00&         0.215&  10.317   &PMS, IR excess, CTTS \\
V16&0.45971 &  67.45303&  18.673&3.167& 13.238&    12.083&     11.605& 1.00 & 0.90&         0.125&  1.137   &PMS \\
V17&0.45217 &  67.51797&  19.933&3.689& 13.794&    12.533&     11.986& 1.00 & 0.45&         0.354&  0.688   &PMS, IR excess, CTTS \\
V18&0.42346 &  67.39261&  13.263&0.884& 11.784&    11.390&     11.331& -    & -   &         0.017&  0.301    &FGMS \\
V19&0.42633 &  67.43320& 14.665&2.420& 10.559&     9.844&      9.496& 1.50 & 3.50 &        0.022,0.014&  0.109,0.101,0.096  &PMS, Herbig Ae/Be \\
V20&0.52329 &  67.45094&  17.483&2.816& 12.653&    11.711&     11.309& 1.50 & 2.00 &        0.095&  0.491    &PMS \\
V21&0.50100 &  67.41972&  12.675&1.258& 10.249&     9.881&      9.731& -    & -    &        0.048&  0.338    &FGMS \\
V22&0.72962 &  67.49997&  18.887&3.531& 12.952&    11.919&     11.558& 0.50 & 0.55 &        0.111&  0.646   &PMS \\
V23&0.58017 &  67.42725& 11.806&2.657&   7.179&     6.552&      6.214& 0.10 & -    &        0.027&  0.217    &PMS \\
V24&0.53228 &  67.40089& 19.595&3.401&  14.298&    13.281&     12.877& 1.50 & 0.65&         0.152&  0.828   &PMS \\
V25&0.58971 &  67.50886& 14.073&1.560&  11.500&    11.175&     10.933&  -   & -    &        0.200&  0.488   &FGMS \\
V26&0.52221 &  67.41542& 18.162&2.938&  13.071&    12.089&     11.723& 2.00 & 1.40 &        0.083&  0.623 &PMS \\
V27&0.60962 &  67.40936& 18.598&3.158&  12.919&    11.864&     11.457& 1.00 & 0.90&         0.085&  1.925    &PMS \\
V28&0.62008 &  67.40633& 20.239&3.555&  13.780&    12.533&     11.793& 1.50 & 0.50&         0.356&  9.735 &PMS, IR excess, CTTS \\
V29&0.46017 &  67.39688& 20.089&2.392&  15.900&    14.416&     14.450&  -   &  -   &        0.300&  1.158    &FGMS \\
V30&0.54121 &  67.37103& 14.151&1.270&  12.032&    11.438&     11.337&  -   &  -   &        0.025&  0.472   &FGMS \\
V31&0.74446 &  67.50539& 19.944&3.485&  14.025&    12.992&     12.457& 1.50 & 0.60&         0.221& 0.516     &PMS, IR excess, CTTS \\ 
V32&0.53858 &  67.40721& 18.703&4.126&  -     &    -      &    -     & 0.10 & 0.30 &        0.203& 0.313     &PMS\\ 
V33&0.53746 &  67.40406& 18.423&3.308&  13.256&    12.296 &    11.939& 0.80 & 0.70&         0.110& 1.014     &PMS \\ 
V34&0.47832 &  67.51580& 19.466&3.209&  13.869&    12.662 &    11.975& 1.50 & 0.70&         0.315& 0.423   &PMS, IR excess, CTTS\\ 
V35&0.29528 &  67.50493& 17.331&2.541&  13.627&    13.246 &    12.994& -    &  -  &        0.030& 0.460     &FGMS \\ 
V36&0.55012 &  67.40846& 17.190&3.229&  12.208&    11.249 &    10.884& 0.30 & 0.90 &        0.072& 0.514    &PMS \\ 
V37&0.46255 &  67.41744& 16.738&2.422&  12.575&    11.718 &    11.400& 10.00& 2.00 &        0.044& 0.796     &PMS \\ 
V38&0.62147 &  67.44061& 17.024&2.783&  12.279&    11.330 &    10.987& 1.00 & 2.50 &        0.064& 1.020     &PMS \\ 
V39&0.59997 &  67.41019& 15.961&2.220&  11.992&    11.169 &    10.896& 10.00& 2.00 &        0.025& 0.325    &PMS  \\ 
V40&0.60189 &  67.42397& 19.467&3.357&  14.388&    14.141 &    13.061& 3.00 & 0.85&         0.175& 1.757     &PMS \\ 
V41&0.41556 &  67.41031& 13.337& 1.221&  11.328&    10.925 &    10.793& -    & -    &       0.029& 0.787     &FGMS  \\
V42& 0.31053&  67.33290& 12.745& 1.134&    10.931&    10.490 &    10.395& -    & -    &     0.021 &0.335    &FGMS \\
\hline
\end{tabular}

${^1}$:foreground MS \\
\end{table*}

In order to find out the possible causes of photometric variations in the identified variable stars in the
region of Be 59, it is important to ascertain their membership.
In the absence of spectroscopic observations, 
we established photometric membership on the basis of $(V,V-I)$ colour-magnitude and
($J-H, H-K)$ colour-colour diagram in next Section.

\section{Cluster Membership}
The colour-magnitude diagram (CMD) of the cluster region clearly reveals contamination due
to field star population (cf. figure 7 by P08). P08 found that the cluster is located at a distance
of 1.00$\pm$0.05 kpc and has a differential reddening $E(B-V)$=1.4-1.8 mag. They also identified a 
foreground field star population at a distance of $\sim$470 pc with a $E(B-V)$=0.40 mag (cf. figure 4
of P08). Therefore, the sample of of identified variables may be contaminated by foreground population. 
We have used $V$, $(V-I)$ CMD and NIR $J-H, H-K$ colour-colour diagram to find out the association of the identified variables with 
the cluster Be 59.
\subsection{$V$, $(V-I)$ Colour-Magnitude Diagram}
Fig. 5 displays $V$, $(V-I)$ CMD for the stars observed in the cluster. The identified variables
are marked by triangles (CTTSs) and filled circles (WTTSs) (cf. section 3.2). The zero-age main-sequence (ZAMS) by Girardi et al. (2002) corrected
for the distance of 1.0 kpc and $E(B-V)$ =1.45 mag is shown by thick continuous line. The pre-main-sequence (PMS)
isochrones for 0.1, 1, 5 and 10 Myrs by Siess et al. (2000) have also been overplotted. The ZAMS corrected for
a distance of 470 pc and $E(B-V)$=0.40 mag for the foreground population is shown by thick dashed curve. Fig. 5.
reveals that the stars V1, V5, V14, V18, V21, V25, V29, V30, V35, V41 and V42 (shown with open circles) follow the ZAMS for field stars, hence could be MS foreground
population.  
It is worthwhile to point that majority of the field stars (10 out of 11) are located outside the core (1.4 arcmin, P08) of the cluster.

In order to determine the ages and masses of the probable PMS variable stars we have used PMS isochrones and evolutionary tracks by Siess et al. (2000), respectively. 
Majority of the probable PMS cluster members have masses in the range 0.3$\lesssim M/M_{\odot}\lesssim$3.5 and ages in the range 1$\lesssim$age$\lesssim$5 Myrs.
This indicates that the young stellar objects associated with the cluster could be TTSs.
Star V23 is the brightest variable in the sample and lies near 0.1 Myr PMS isochrone.

\subsection{NIR Colour-Colour Diagram}
The NIR colour-colour diagram (CCD) can be used to distinguish the CTTSs and WTTSs. Fig. 6 shows NIR CCD
for the identified variables. The NIR data have been taken from the 2MASS Point Source Catalogues (PSC; Cutri et al. 2003). The 2MASS counterpart of the variables were searched using a match value of ~3 arcsec. 
Of the 42 detected variables, we found 2MASS counterparts for 41 stars (30 cluster members and 11 foreground MS stars). 
The 2MASS counterparts for V32 star could not be found. It is the reddest variable star in the sample and lying above 0.1 Myr PMS isochrone in $V, (V-I)$ CMD. However, on the basis of the location of this star on $V, (V-I)$ CMD, we have 
considered V32 as a PMS object.
The 2MASS magnitudes and colours were transformed to the CIT system using the relations given on the 2MASS website. The solid
and long dashed lines represent unreddened MS and giant branch (Bessell \& Brett 1988), respectively.
The dotted line indicates intrinsic locus of CTTSs (Meyer et al. 1997). 
The parallel dashed lines are the reddening vectors drawn from the tip (spectral type M4) of the giant branch (`upper reddening line'), from the base (spectral type A0) of the main-sequence branch (`middle reddening line') and from the tip of the intrinsic CTTS line (`lower reddening line'). The extinction ratios $A_J/A_V= 0.265, A_H/A_V= 0.155$ and $A_K/A_V= 0.090$ have been adopted from Cohen et al. (1981).  
We classified the sources into three regions in the NIR CCD (cf. Ojha et al. 2004a). The `F' sources are those that are located between 
the upper and middle reddening lines and are considered to be either field stars (main-sequence stars, giants) or Class III and Class II sources 
with small NIR excesses.
The identification of Class II sources having a small amount of NIR excess is difficult on the basis of ($J-H, H-K)$ CCD. However, recent 
studies, where Class II sources were identified on the basis of mid-IR observations through Spitzer telescope, manifest that these Class II objects lie in 
`F' region near the middle reddening vector (e.g. Chauhan et al. 2011).
 The `T' sources are those that are located between the middle and lower reddening lines. These sources are considered to be 
mostly CTTSs (Class II objects).
`P' sources are those that are located in the region redward of the ‘T’ region and are most likely Class I objects (protostar-like objects; Ojha et al. 2004b). 

Majority of the stars lying in the `F' region are considered as Class III stars/WTTSs and are shown by filled circles.
However, there seems segregation of sources between $(J-H)$ $\gtrsim$ 1.0 mag and $(H-K)$ $\gtrsim$ 0.50 mag in $(J-H, H-K)$ colour-colour diagram.
As discussed above that Class II sources with small amount of NIR excess could be found in `F' region near the middle reddening
vector. 
The stars having $(J-H)$ $\gtrsim$ 1.0 mag and $(H-K)$ $\gtrsim$ 0.50 mag are considered
CTTS and these are shown by filled triangles.
The foreground MS stars as identified from $V, V-I$ CMD are distributed
around the ZAMS and shown by open circles. The nature of the identified variables is mentioned in the Table 2. 

The membership of the variable stars is determined on the basis of their location in $(V, V-I)$ CMD and NIR CCD. 
However, spectroscopic follow up of these objects is needed to confirm their nature. 

\section{Period Determination}
We used the Lomb-Scargle (LS) periodogram (Lomb 1976; Scargle 1982) to determine the most likely
period of a variable star. The LS method is useful for finding 
significant periodicities even with unevenly sampled data. We used the algorithm publicly available
at the Starlink\footnote[2]{http://www.starlink.uk} software database,
and verified the periods further with the software period04
\footnote[3]{http://www.univie.ac.at/tops/Period04} (Lenz \& Breger 2005). 
The software period04 provides the frequency and semi-amplitude of the variability in a light curve.
Periods derived from the LS method and Period04 generally
matched well. In order to further verify the period estimates we also used periodogram analysis
available online\footnote[4]{http://nsted.ipac.caltech.edu/periodogram/cgi-bin/Periodogram/nph-simpleupload}.  
For any star showing a spurious period, we visually inspected the phased light curve for that particular period.
The phased light curves for CTTSs, WTTSs and field stars are shown in Figs. 7, 8 and 9, and 10 respectively.  
Figs. 7, 8, 9 and 10 plot the mean of differential magnitudes in 0.04 phase bin as a function of phase. The error bars represent the $\sigma$ of the mean.  
In subsequent section we discuss the variability characteristics of individual variable star. 

\begin{figure*}
\includegraphics[width=15cm,height=10cm]{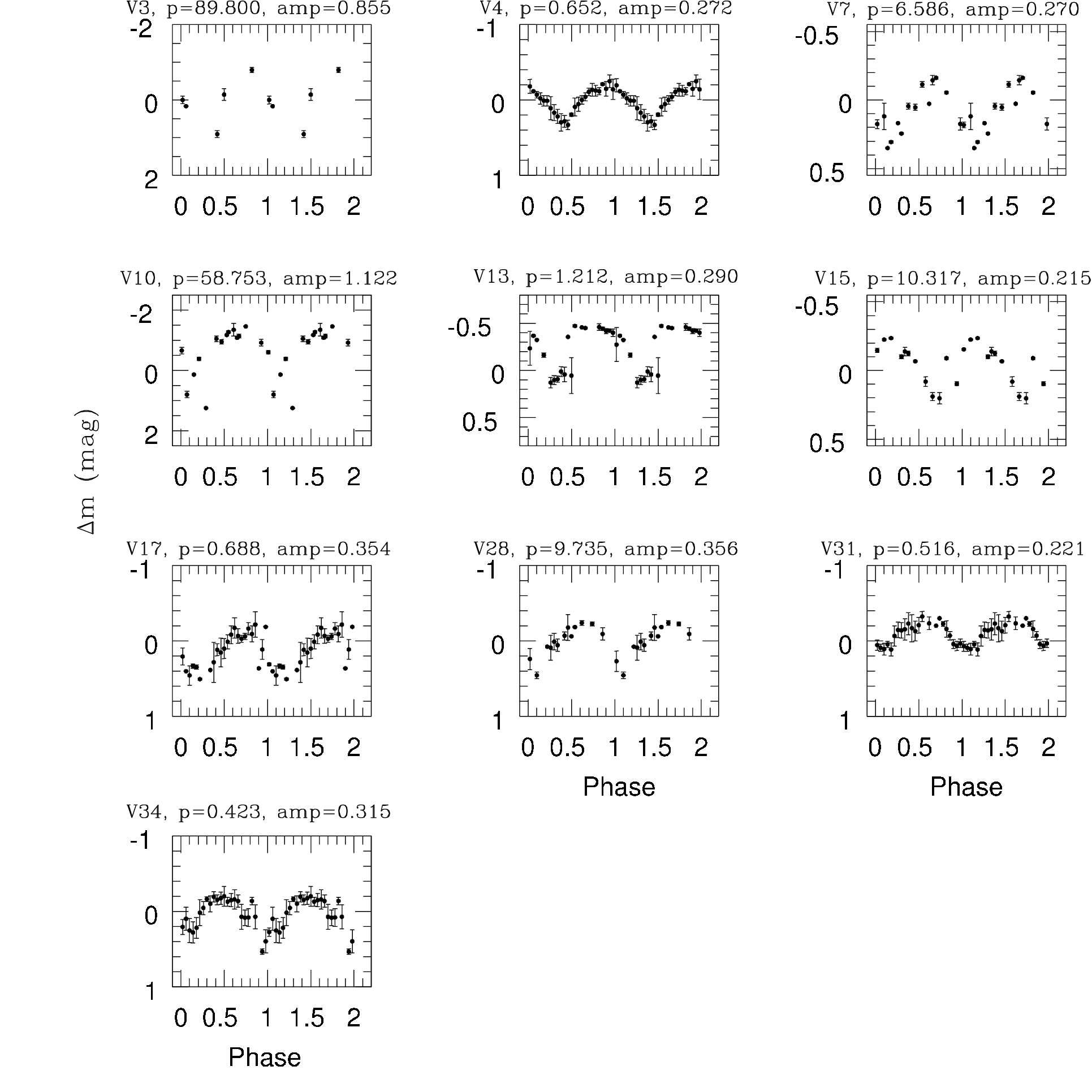}
\caption{The phased light curves of probable CTTSs candidate variables identified in Be 59.}
\end{figure*}

\begin{figure*}
\includegraphics[width=15cm,height=10cm]{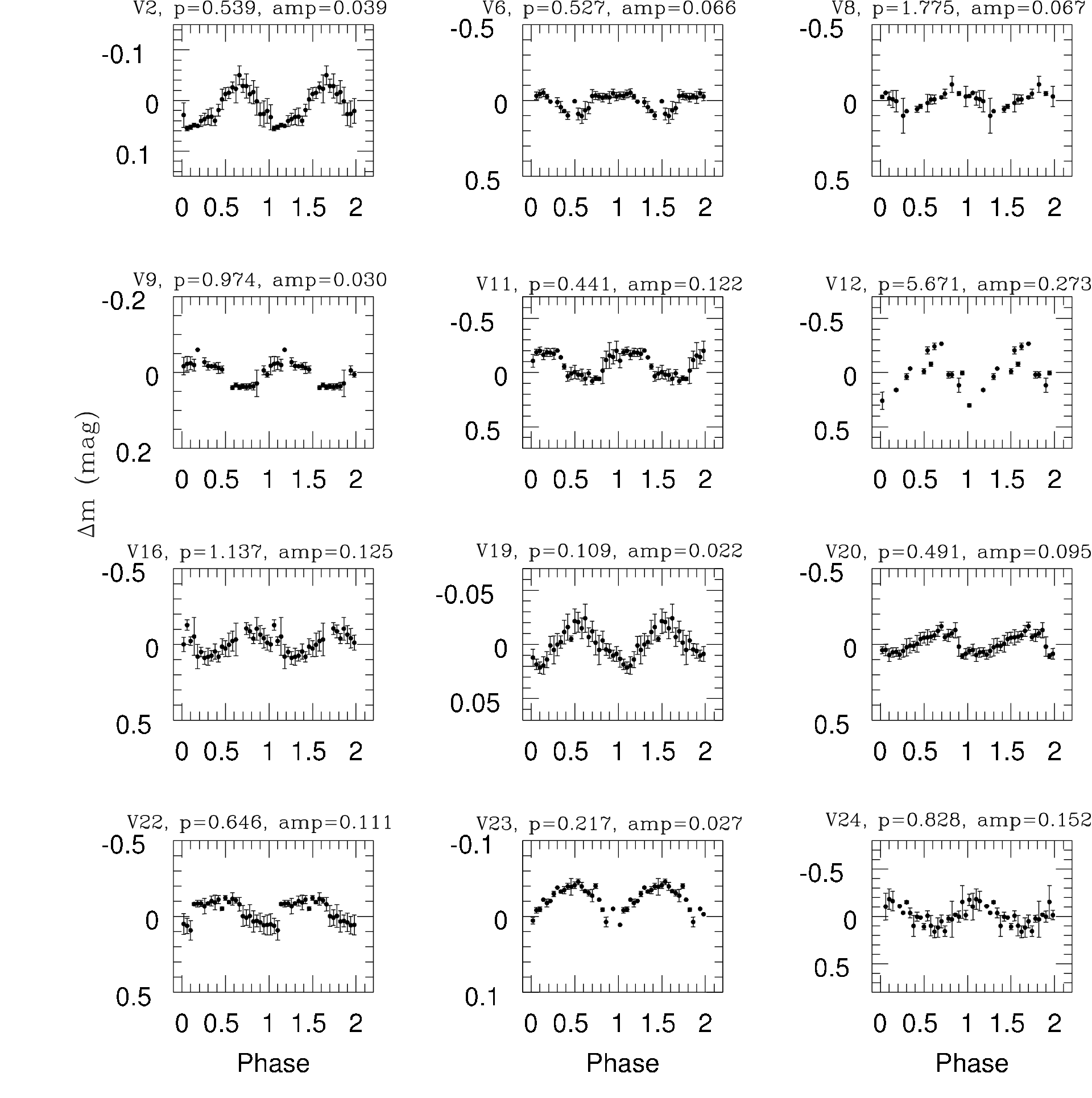}
\caption{The phased light curves of probable WTTSs candidate variables identified in Be 59.}
\end{figure*}

\begin{figure*}
\includegraphics[width=15cm,height=10cm]{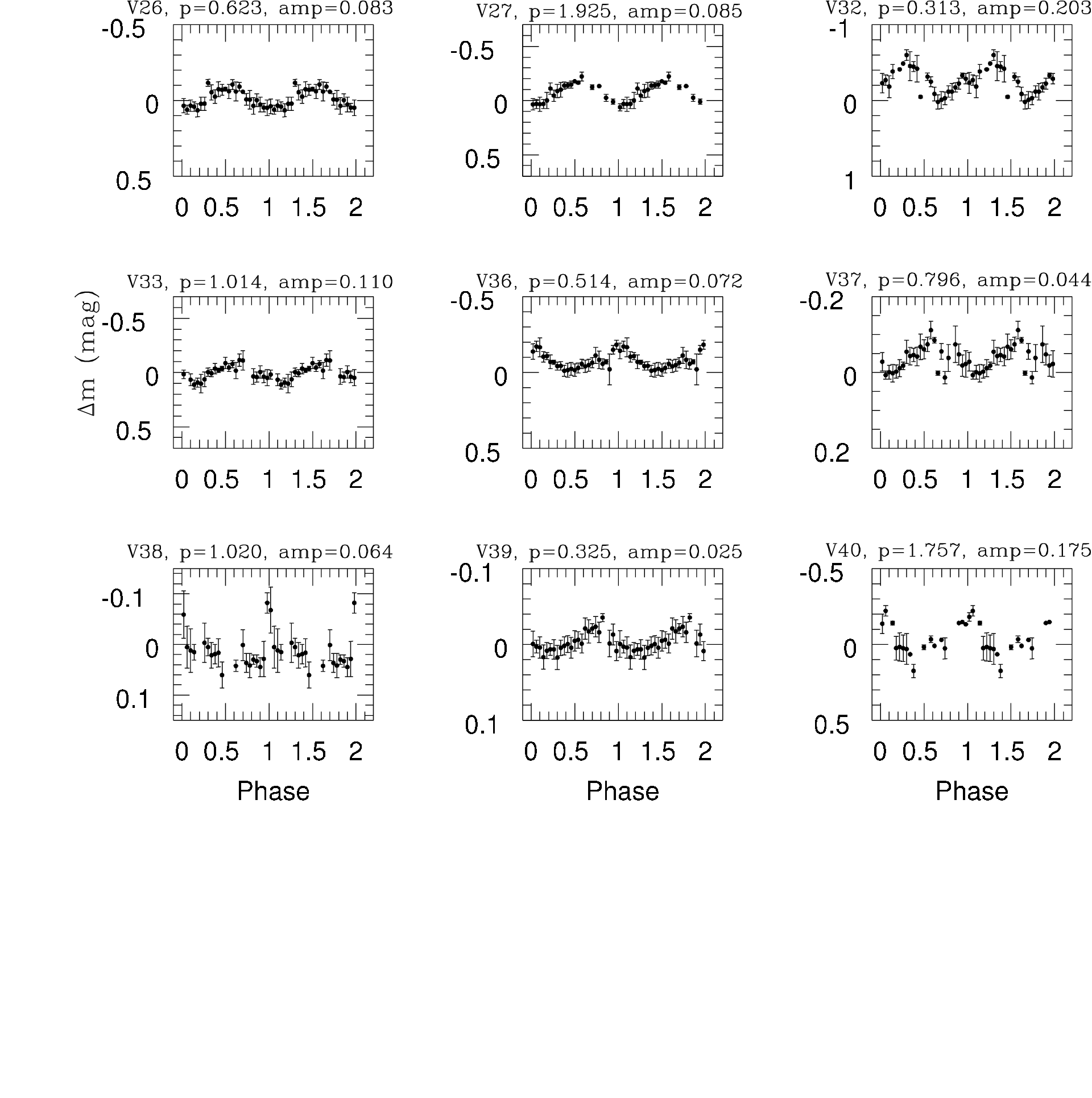}
\caption{Same as Fig. 8.}
\end{figure*}

\begin{figure*}
\includegraphics[width=15cm,height=10cm]{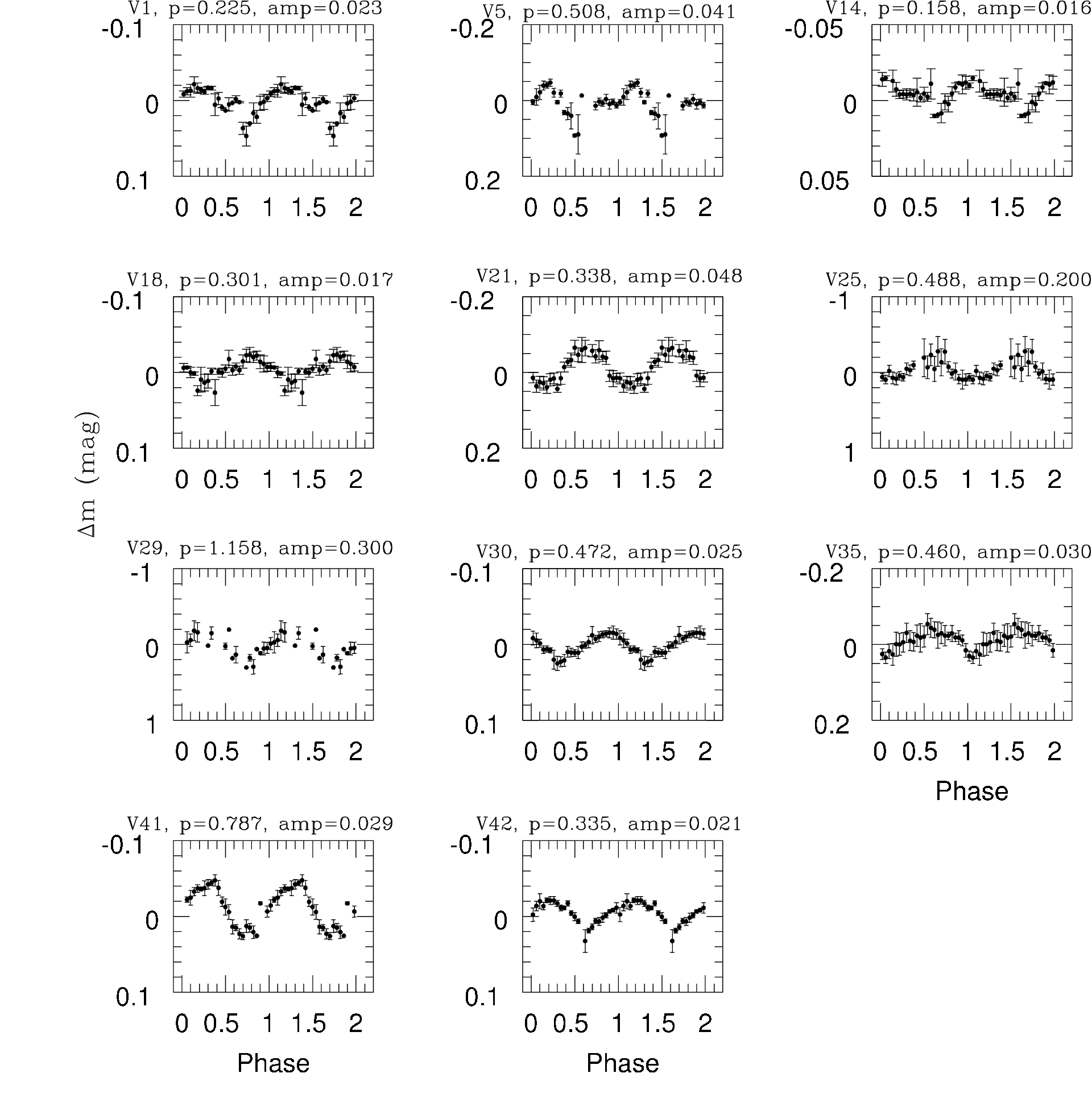}
\caption{The phased light curves of foreground MS variables.}
\end{figure*}


\section{Variability Characteristics}
\subsection{Non Periodic Variables}
Light curve of one WTTS variable named V38 displayed in Fig. 9 does not show periodic variability.
This star could be either irregular variable or 
the period estimated in the present work could be in error.

\subsection{Periodic Variables}
The short term periodic variations (2 $\sim$ 10 days) in TTSs are believed to occur 
due to the axial rotation of a star with an inhomogeneous surface, having either hot or cool spots (Herbst et al. 1987,
1994).
Herbst et al. (1994) identified three types of TT variables based on their variability timescales ranging from a
day to weeks. Type I variability,
most often seen in WTTSs, is characterized by smaller stellar flux variations (a few times of 0.1 mag) and results from the rotation of a cool spotted photosphere. Type II variations have larger brightness variations
(up to $\sim$2 mag), most often irregular but sometime periodic, are associated with short-lived accretion related hot spots at the stellar surface of CTTSs. The rare type III variations are characterized by luminosity dips lasting from a 
few days upto several months, which presumably result from circumstellar dust obscuration. The sample of TTSs in the
cluster Be 59 indicates that $\sim$90\% TTSs have periods less than 15 days. 
The estimated periods of 8 CTTSs vary in the range of 0.652 day to 10.317 day, however stars V3 and V10 have significantly larger
periods (89.8 day and 58.7 day respectively). 
The period estimates for WTTSs vary from 0.109 day to 5.671 day,  
however $\sim$95\% WTTSs have periods $\le$ 2 day.
The shorter periods in the case of WTTSs could be explained as
most of the stars at this age might have unlocked from their disks while contracting towards MS. They spin up because of decreasing
radii in order to conserve their angular momentum (Strom et al. 1989; Skrutskie et al. 1990; Haisch et al. 2001; Bouvier et al. 1994; Herbst \& Mundt 2005; Lamm et al. 2005; Kundurthy et al. 2006). But Nordhagen et al. (2006) have different view and suggest that environment also plays a crucial role in establishing the rotation periods besides age and mass. The presence of `O' type stars in the vicinity would speed up the process of the dispersal of the circumstellar disks causing the stars to 
spin up at a much younger age. 

The amplitude of WTTSs ranges from 0.022 to 0.273 mag. 
The amplitude in the case of CTTSs has a range of 0.215 
to 1.122 mag. 
The present results 
 regarding amplitudes are in agreement with Grankin et al. (2008 for WTTSs)
and Grankin et al. (2007 for CTTSs).
The brightness of CTTSs is found to vary with larger amplitude in comparison to the WTTSs. The large amplitude in the case of CTTSs indicates   
presence of the hot spots on their surfaces. The smaller amplitude in the case of WTTSs suggests that
these PMS stars might be in the process of dissipation of their circumstellar disk.

The star V3, a probable CTTS, seems to be an interesting object. It has relatively high excess ($H-K=1.23$) and shows relatively high extinction (cf. Fig. 6). It has a period of 89.80 days with amplitude variation of 1.0 mag and 
its mass is estimated to be $\sim$0.9 $M_{\odot}$. The relatively high  
extinction of this object could be due to the presence of the circumstellar disk. Edwards et al. (1993)
found that the stars which showed the evidence of circumstellar disks, inferred from their $(H-K)$ colours, are slow
rotators in comparison to those that lacked circumstellar disks.
Relatively high extinction and longer period of V3 support this argument.
This star may belong to type III as classified by Herbst et al. (1994), where
relatively larger luminosity dips for longer period are expected due to circumstellar dust obscuration.

The H$\alpha$ emission star V10, classified as CTTS,  also seems to be an interesting object. A careful inspection of
night to night observations of this star reveals amplitude variation. Its 2006 observations show that the brightness of this star was varying with larger 
amplitude (1.122 mag) and period 1.102 day. From 2007 and onwards, it shows consistent brightness variation with amplitude of $\sim$0.20 mag and 
period of 1.102 day. Whereas, the whole data set yield a period of 58.753 day. 

Star V19 (($V=14.665$ mag) could be a PMS pulsating star as suggested by its variability characteristics like period, amplitude and
the shape of light curve. The period and amplitude of star V19 vary over time. Its period
on 11 October 2009 and 27 October 2010 is found to be 0.101 day and 0.096 day, respectively, whereas its amplitude varied from 0.022 to 0.014 mag on 
respective dates. Considering all the observations of this star its period comes out be 0.109 day.
Star V19 might be a $\delta$ Scuti type variable star.

An early type star V23 (brightest star in the present sample), considered to be a member of the cluster, might be a pulsating star.
Its period (0.217 day) and amplitude (0.027 mag) are very much similar to the B-type pulsating stars.

Star V21, lying just below the MS, could be a cluster member.
The period, amplitude and the shape of light curve of V21 suggest that it might be a pulsating star.
The spectroscopic observations of star V21 was made by Majaess et al. (2008). They considered it to be a member of the cluster Be 59 and found it to a be B-type star.

The field variables have period $<$ 1.16 days and amplitude $<$ 0.30 mag.

\subsection{Period and Amplitude Distribution}
In order to study the correlation between stellar activity and mass/age, we plot rotation period 
as a function of mass/age in Fig. 11.
The
rotation periods of TTSs associated with the cluster Be 59 do not show any dependency on the mass (Fig. 11 left
panel).
Fig.11 (right panel) shows two long periodic variables (V3 with P=89.8 day and V10 with P=58.753 day, cf. Table 2) have ages less than 5 Myrs, which 
could be interpreted as the rotation is regulated by disk locking at early ages (Jayawardhana et al. 2006, Herbst et al. 2002). The presence of NIR excess in 
$(H-K)$ of V3 and H$\alpha$ emission in case of V10 is probably indicative of disk locking. But it is worth to mention here that large IR excess alone could not indicate 
ongoing disk-locking because IR excess is the indicator for the presence of circumstellar dust and  it is not an indicator for current mass accretion
(Lamm et al. 2005).

Fig. 12 (left panel) and Fig. 12 (right panel) show
amplitude as function of mass and age, respectively. 
Fig. 12 (left panel) reveals
that the amplitude of TTSs seems to be correlated with the mass of the star in the sense that
amplitude decreases with increasing mass of the variables. The largest amplitude is found 
in the case of V3 (0.855 mag) and V10 (1.122 mag). The star V3 shows significantly higher $(H-K)$ excess with 
relatively larger extinction, whereas star V10 shows H$\alpha$ emission. Thus both the stars manifest presence 
of disk. The decrease in amplitude could be due to the dispersal of the disk material. If this notion is true, it indicates that the mechanism for disk dispersal operates less efficiently for relatively low
mass stars. In fact there is evidence in literature that disk dispersal mechanism depends on the mass of the 
stars (e.g. Carpenter et al. 2006).
But in Fig 12. (right panel), the amplitude variation with ages shows no definite trend, and majority of the sources having the ages within 5 Myrs show the amplitude of variability up to 0.35 mag with two CTTSs as exception, which have been discussed earlier.

\section{Summary}
The paper presents our efforts to search for variable stars in young 
open cluster Be 59. Using $VI$ photometry of a $\sim13^{'}\times13^{'}$ field around the cluster
we have identified 42 variable stars.  
The probable members of the cluster are identified using $(V, V-I)$ colour-magnitude diagram and $(J-H, H-K)$ colour-colour 
diagram.
Out of 42 variables, 31 variable stars are found to be the probable PMS stars associated with the cluster Be 59. 
The majority of the probable PMS candidate variable stars have ages 1 to 5 Myrs.
The masses of probable PMS variable stars range from $\sim$0.3 to $\sim$3.5 M$_{\odot}$, suggesting that these could be TTSs.
The sample of TTSs in the
cluster Be 59 indicates that $\sim$90\% TTSs have periods less than 15 days.
The CTTSs have larger amplitudes in comparison to WTTSs.
The larger amplitude of CTTSs indicates the
presence of hot spots on their surface. The smaller amplitude in the case of WTTSs suggests that
these PMS stars might be in the process of dissipation of their circumstellar disk.
It is found that the amplitude of TTSs decreases with increasing mass of the variable stars.
The decrease in amplitude of variability of TTSs could be due to the dispersal of the disk material, indicating that the mechanism for 
disk dispersal operates less efficiently for relatively low mass stars.

\begin{figure}
\includegraphics[width=8cm]{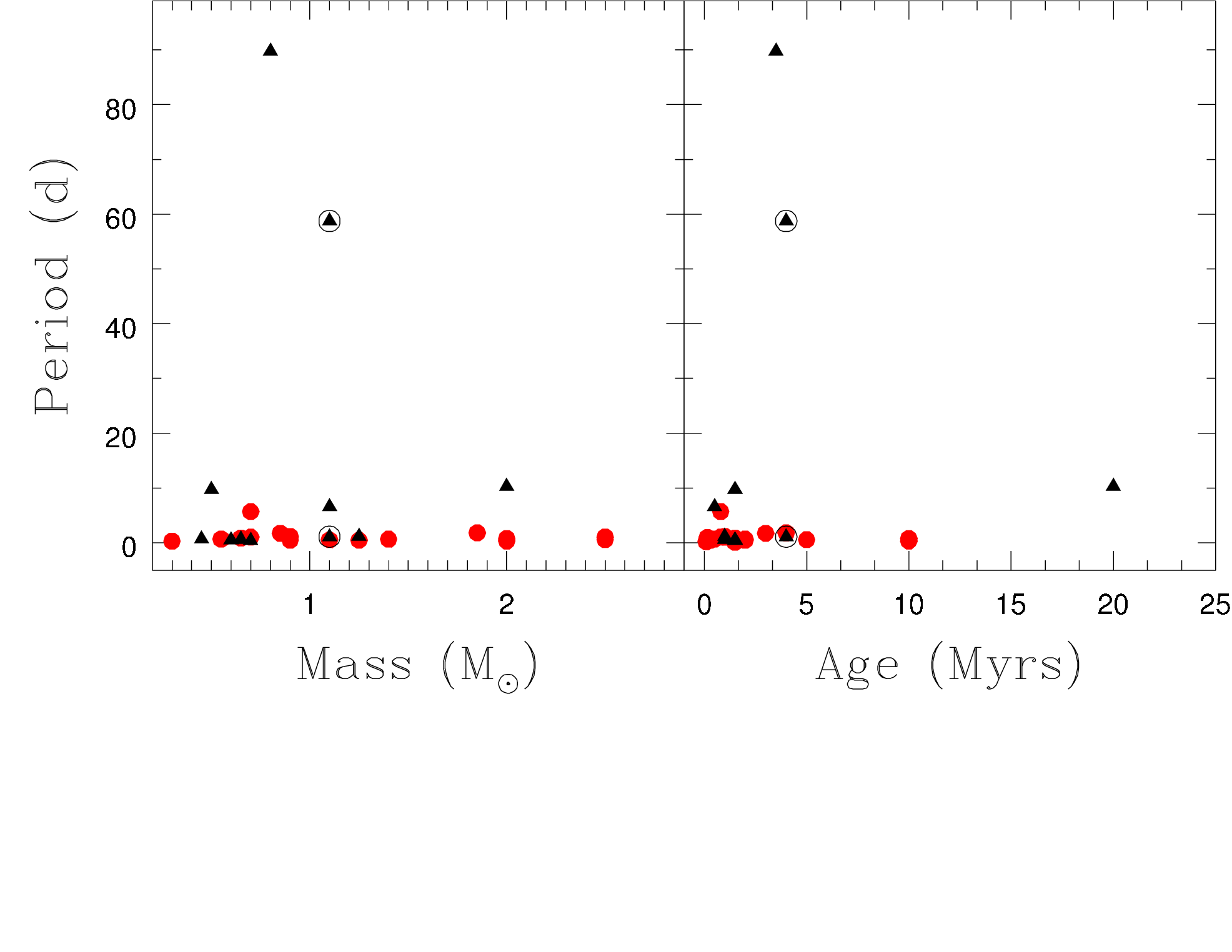}
\caption{Rotation period as a function of mass and age of CTTSs and WTTSs. The symbols are same as in Fig. 5.
The encircled triangles represent long and short periods for V10.}
\end{figure}

\begin{figure}
\includegraphics[width=8cm]{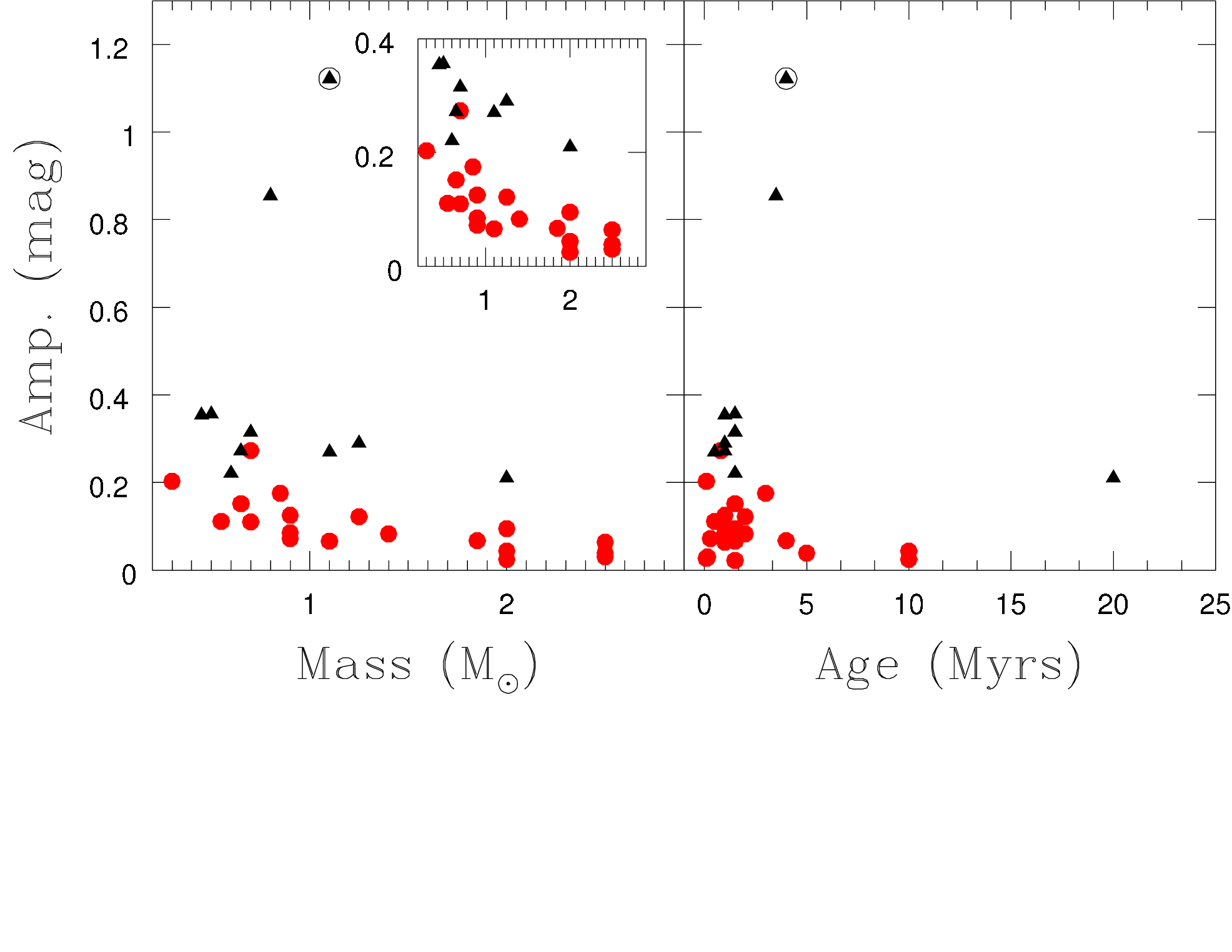}
\caption{Amplitude as a function of mass and age of CTTSs and WTTSs.  The symbols are same as in Fig. 5 and Fig. 11. Inset
in the left panel shows a magnified view of amplitude variation.} 
\end{figure}

\section{Acknowledgments}
Authors are very thankful to the anonymous referee for useful suggestions, which improved the scientific contents of the paper.

\label{lastpage}

\end{document}